\documentclass[aps,prd,twocolumn,nofootinbib,floatfix,showpacs,showkeys,preprintnumbers]{revtex4}
\usepackage{subfigure}
\usepackage{graphicx}
\usepackage{dcolumn}
\usepackage{epsfig}
\usepackage{amsmath}
\usepackage{amsfonts}
\usepackage{amssymb}
\usepackage{color}
\usepackage{hyperref}
\usepackage{ulem}

\usepackage{hhline}
\usepackage{array}
\newcommand{\PreserveBackslash}[1]{\let\temp=\\#1\let\\=\temp}
\newcolumntype{C}[1]{>{\PreserveBackslash\centering}p{#1}}
\newcolumntype{R}[1]{>{\PreserveBackslash\raggedleft}p{#1}}
\newcolumntype{L}[1]{>{\PreserveBackslash\raggedright}p{#1}}

\newcommand{\be}{\begin{equation}}
\newcommand{\ee}{\end{equation}}
\newcommand{\bea}{\begin{eqnarray}}
\newcommand{\eea}{\end{eqnarray}}
\newcommand{\nn}{\nonumber}
\newcommand{\di}{\text{d}}

\newcommand{\bz}{{\bf z}}

\newcommand{\bk}{{\bf k}}

\begin{document}

\title{Cosmic distance inference from purely geometric BAO methods:\\
Linear Point standard ruler and Correlation Function Model Fitting}
%\title{Cosmological-distance inference from Purely-Geometric-BAO:\\ 
%the Linear Point standard ruler and correlation-function model fitting}

\author{Stefano Anselmi}
\email{stefano.anselmi@iap.fr}
%\affiliation{Department of Physics/CERCA/Institute for the Science of Origins, Case Western Reserve University, Cleveland, OH 44106-7079 -- USA}
\affiliation{LUTH, UMR 8102 CNRS, Observatoire de Paris, PSL Research University, Universit\'e Paris Diderot, 92190 Meudon -- France}
\affiliation{Institut d'Astrophysique de Paris, CNRS UMR 7095 and UPMC, 98bis, bd Arago, F-75014 Paris -- France}

\author{Pier-Stefano Corasaniti}
\affiliation{LUTH, UMR 8102 CNRS, Observatoire de Paris, PSL Research University, Universit\'e Paris Diderot, 92190 Meudon -- France}

\author{Ariel G.~Sanchez}
\affiliation{Max-Planck-Institut f\"ur extraterrestrische Physik, Postfach 1312, Giessenbachstr., 85741 Garching -- Germany}

\author{Glenn D.~Starkman}
\affiliation{Department of Physics/CERCA/Institute for the Science of Origins, Case Western Reserve University, Cleveland, OH 44106-7079 -- USA}

\author{Ravi K.~Sheth}
\affiliation{Center for Particle Cosmology, University of Pennsylvania, 209 S. 33rd St., Philadelphia, PA 19104 -- USA}
\affiliation{The Abdus Salam International Center for Theoretical Physics, Strada Costiera, 11, Trieste 34151 -- Italy}

\author{Idit Zehavi}
\affiliation{Department of Physics/CERCA/Institute for the Science of Origins, Case Western Reserve University, Cleveland, OH 44106-7079 --- USA}

\date{\today}

\begin{abstract}
Leveraging the Baryon Acoustic Oscillations (BAO) feature present in clustering 2-point statistics, we aim to measure cosmological distances independently of the underlying background cosmological model. However this inference is complicated by late-time non-linearities that introduce model and tracer dependencies in the clustering correlation function and power spectrum, which must be properly accounted for. With this in mind, we introduce the ``Purely-Geometric-BAO,'' which provides a rigorous tool to measure cosmological distances without assuming a specific background cosmology. We focus on the 2-point clustering correlation function monopole, and show how to implement such an inference scheme employing two different methodologies: the Linear Point standard ruler (LP) and  correlation-function model-fitting (CF-MF). For the first time we demonstrate how, by means of the CF-MF, we can measure very precisely  the sound-horizon/isotropic-volume-distance ratio, $r_{d}/D_{V}(\bar{z})$, while correctly propagating all the uncertainties. Using synthetic data, we compare the outcomes of the two methodologies, and find that the LP provides up to $50\%$ more precise measurements than the CF-MF. Finally, we test a procedure widely employed in BAO analyses: fitting the 2-point function while fixing  the cosmological and the non-linear-damping parameters at fiducial values. We find that this underestimates the distance errors by nearly a factor of $2$. We thus recommend that this practice be reconsidered, whether for parameter determination or model selection.
\end{abstract}

\pacs{}
\keywords{large-scale structure of Universe}

\maketitle

%\newcommand{\ste}[1]{\textcolor{red}{\textbf{\small[Ste: #1]}}}

%%%%%%%%%%%%%%%%%%%%%%%%%%%%%%%%%%%%%%%%%%%%%%%%%%%%%%%%%%%%
%%%%%%%%%%%%%%%%%%%%%%%%%%%%%%%%%%%%%%%%%%%%%%%%%%%%%%%%%%%%
\section{Introduction}
\label{intro}

In the primordial photon-baryon plasma, the opposing effects of gravity and thermal pressure generated acoustic waves that propagated until decoupling, leaving an imprint of the scale of the sound horizon on the distribution of matter in the universe.
This process caused the Baryon Acoustic Oscillations (BAO), which manifest as ripples in the clustering power spectrum (PS) and as an {\it acoustic} peak in the 2-point correlation function (CF) \cite{2005MNRAS.362..505C, Eisenstein:2005su}.

This characteristic imprint of the BAO on the CF was recognized long ago 
as a powerful standard ruler, which could be used to map the expansion history of the Universe.  
This has motivated a major effort in the design and realization of a new generation of galaxy surveys such as Euclid,\footnote{\url{http://sci.esa.int/euclid/}} DESI,\footnote{\url{http://desi.lbl.gov}} and WFIRST\footnote{\url{https://wfirst.gsfc.nasa.gov}}. 
The underlying idea is that the comoving length of this ruler is independent of the cause of the late-time acceleration of the Universe, its spatial-geometry, the primordial fluctuation parameters and the observed tracers of the matter-density field (e.g.~galaxies). 

Endowed with these standard-ruler properties, the BAO are ideal for exploiting the Alcock-Paczynski distortions and inferring cosmic distance measurements (though see \cite{2013MNRAS.434.2008S, 2011MNRAS.411..277S, 2018arXiv180810695M} for alternative approaches). These distortions can be modeled analytically, parametrised, and measured from survey data, to finally provide an estimate of the Hubble parameter and the angular-diameter distance \cite{2013MNRAS.431.2834X}.

It is common understanding that the position of the acoustic peak in the CF is a geometric standard ruler. In linear theory this is indeed the case, since the peak position satisfies the required properties. 
However, due to late-time non-linearities that affect in different ways the clustering of the matter field and that of its tracers \cite{2016MNRAS.458.1909V, 2010PhRvD..82j3529D, 2017PhRvD..95d3535B, 1986ApJ...304...15B}, the peak position is time-dependent and cannot be simply deployed as a standard ruler \cite{2008PhRvD..77d3525S, 2008MNRAS.390.1470S}.

To overcome these limitations, current standard analyses employ a fitting approach usually called {\it BAO-Only} (BAO-O) \cite{2008ApJ...686...13S, 2012MNRAS.427.2146X, 2014MNRAS.441...24A}. This relies on fitting a template function of the CF, parametrized in term of the linear CF, a damping term and broad-band nuisance parameters. In fitting the CF data, a set of fiducial cosmological parameter values is assumed to fix the predicted linear CF. Similarly the damping term is set to a fiducial value, while the broad-band parameters are marginalized over. The latter step has the intent of propagating theoretical and observational systematics that could otherwise bias the analysis.

While simulations and survey mock catalogs are used to show that the distances estimated using BAO-O are unbiased, their precise information content is unclear. Fixing the  cosmological parameters to precise fiducial values adds  information that we cannot quantify. This could cause the distance errors to be underestimated because the uncertainty in them is neglected.
Meanwhile, the value of the time-dependent damping parameter is derived through galaxy-survey-mocks, run for the fiducial cosmology. Since we do not have {\it ab-initio} galaxy mocks, and the value of the damping parameter is also tracer dependent \cite{2016MNRAS.458.1909V, 2010PhRvD..82j3529D, 2017PhRvD..95d3535B, 1986ApJ...304...15B}, fixing it exactly ascribes to it unjustified precision and accuracy, potentially with similar effects on the distance estimates. All these assumptions may lead to underestimated distance errors. A final point of concern is that mock catalogs are generated from flat-$\Lambda$CDM model simulations, and the inferred BAO-O errors cannot be naively extrapolated to non-standard cosmological scenarios. Because of this, cosmological parameter constraints inferred in combination with BAO-O distance estimates should  be limited to the parameter space of the flat-$\Lambda$CDM scenario, though there too a systematic effect on the inferred bounds cannot be {\it a priori} excluded. However, the applicability of these estimates to the parameter space of Dark Energy (DE) and non-flat models, such as the DE equation-of-state parameter or the spatial curvature, demands even more precaution. Used in such a context, as for instance in \cite{2015PhRvD..92l3516A, 2017NatAs...1..627Z, 2018JCAP...05..033H}, the unpropagated uncertainties on BAO-O distance measurements can potentially lead to erroneous conclusions.

In this work, we present a rigorous definition of what we refer to as {\it Purely-Geometric-BAO} (PG-BAO) methods, which restore the original standard-ruler role of the BAO. PG-BAO methods do not assume spatial flatness or a specific model for the late-time acceleration, do not require accurate knowledge of the dark-matter/tracer relation, and are independent of the primordial-fluctuation parameters. This can be achieved either
\begin{enumerate}
\item[(A)] by identifying a feature in the BAO range of scales that is independent of the primordial-fluctuation parameters, redshift, and tracer, and that can therefore be used to infer distance estimates;
\item[(B)] by performing a model-fitting analysis of the CF to infer a cosmic distance estimate, while carefully marginalising over the primordial-fluctuation parameters and any other parameters that change with redshift %\footnote{For some particular cases some time dependent parameters are not DE dependent, e.g.~the growth rate in standard quintessence models with constant DE equation-of-state parameter. However this is not relevant as the growth rate in BAO studies does never appear as a parameter and, in addition, we aim at targeting a wider set of DE models.} 
or the choice of target tracer.
\end{enumerate}

In \cite{2016MNRAS.455.2474A}, the authors have shown that the Linear Point standard ruler (LP) provides a PG-BAO method, as it precisely shares all the (A) requirements. It is defined as the mid-point between the BAO peak and the dip of the 2-point correlation function. It is insensitive to non-linear gravity, redshift-space distortions and scale-dependent bias at the $0.5\%$ level. Moreover, perturbation-theory arguments demonstrate that its properties hold when smooth DE models are considered \cite{Anselmi:2014nya}. It has been successfully validated against mock galaxy-clustering data, and employed to measure distances from galaxy surveys in a model-independent way \cite{2018PhRvD..98b3527A, 2018PhRvL.121b1302A}. 

Here, for the first time, we introduce a procedure which provides a PG-BAO distance estimates that satisfies the (B) requirements. Hereafter, we will refer to this procedure as CF-Model-Fitting (CF-MF) analysis. %(B) can be implemented by applying it to a CF-model. 
We stress that, to properly account for the parameter dependencies, we need to fit CF data in Mpc and not in Mpc/h units. We identify the quantities that are sensitive to the nuisance information and we marginalize over them. We show that cosmic distances are poorly constrained by this procedure. However, the sound-horizon scale embedded in the CF provides a standard ruler that can be indirectly extracted from the CF-MF analysis. Quite remarkably, we find that the ratio of the CF-MF inferred sound-horizon scale to the cosmic distance is determined with up to {\it 50 times better precision}.

We compare the results of LP versus CF-MF measurements, and show that the LP standard ruler provides up to $50\%$ smaller errors compared to the CF-MF approach. We also test the impact of the assumptions employed in the BAO-O method, namely fixing the cosmological and damping parameters to fiducial values while marginalizing over broad-band nuisance parameters. We find that, when compared to the CF-MF approach, these assumptions lead to distance uncertainties being underestimated by nearly a factor of $2$. Hence, we advocate to reconsider these assumptions in BAO analyses that aim to provide model-independent distance measurements.

The paper is organized as follows. In Section \ref{sec:method} we first define PG-BAO methods. For practical purposes we introduce a CF synthetic data model and its covariance, then we explain how the Alcock-Paczynski equation can be exploited to extract distance information through both the LP and CF-MF approaches. In Section \ref{sec:results} we detail the survey characteristics and discuss the results of the distance errors inferred from LP and CF-MF respectively. Finally, in Section \ref{sec:concl} we present our conclusions. %We provide details on the procedure to numerically stabilize the Fisher-matrix results in Appendix \ref{appendix:Fisher}.

%%%%%%%%%%%%%%%%%%%%%%%%%%%%%%%%%%%%%%%%%%%%%%%%%%%%%%%%%%%%
%%%%%%%%%%%%%%%%%%%%%%%%%%%%%%%%%%%%%%%%%%%%%%%%%%%%%%%%%%%%

%%%%%%%%%%%%%%%%%%%%%%%%%%%%%%%%%%%%%%%%%%%%%%%%%%%%%%%%%%%%%%%%%%%%%%%%%%%%%%%%%%%%
%%%%%%%%%%%%%%%%%%%%%%%%%%%%%%%%%%%%%%%%%%%%%%%%%%%%%%%%%%%%%%%%%%%%%%%%%%%%%%%%%%%%

\section{Methodology}\label{sec:method}
A purely geometric approach to BAO (PG-BAO) consists in inferring distance measurements that exploit the BAO feature from galaxy-survey observations without the need to assume a $\Lambda$CDM cosmology, a specific DE model, or a flat geometry. In such a case, the distance measurements can be used to constrain the curvature of the universe or the time-evolution of DE. This can be achieved if the functional form of the CF in the BAO range of scales is model-independent. This is indeed the case for the $\Lambda$CDM scenario and for smooth standard/clustering quintessence models. However, we will see that an optimal PG-BAO distance measure requires only a DE-independent comoving standard ruler. For this purpose we introduce a synthetic CF-data model, which provides us with the setup to compare distance errors inferred from the LP standard ruler and the CF-MF approach. In the LP case, cosmic distances are estimated through a model-independent parametric fit of the synthetic CF-data using a polynomial fitting function (as described in \cite{2018PhRvL.121b1302A}). Then, statistical and systematic errors for the LP approach are inferred by propagating the LP errors from the uncertainties on fit polynomial coefficients \cite{2018PhRvD..98b3527A}. In the CF-MF case, we instead estimate distance errors through a Fisher-matrix analysis. This considerably reduces the error-analysis computational cost, but provides optimal errors. The comparison between LP-derived uncertainties and those from the CF-MF will prove to be conservative, as we  compare realistic LP errors versus optimal CF-MF ones.

 %%%%%%%%%%%%%%%%%%%%%%%%%%%%%%%%%%%%%%%%%%%%%%%%%%%%%%%%%%%%%%%%%%%%%%%%%%%%%%%%%%%%
%%%%%%%%%%%%%%%%%%%%%%%%%%%%%%%%%%%%%%%%%%%%%%%%%%%%%%%%%%%%%%%%%%%%%%%%%%%%%%%%%%%%
\subsection{Synthetic Data Model}
\label{sec:synthetic}

%%%%%%%%%%%%%%%%%%%%%%%%%%%%%%%%%%%%%%%%%%%%%%%%%%%%%%%%%%%%%%%%%%%%%%%%%%%%%%%%%%%%
\subsubsection{Non-Linear Correlation Function Model}
\label{sec:CFmodel}

We adopt a simple but accurate analytic approximation to the non-linear CF in the BAO range of scales. This allows us to know precisely the position of the ``true'' LP and compare it with the estimated position, and to use the Fisher-matrix formalism for the evaluation of the CF-MF distance errors, while keeping under control the parameter dependence, disentangling the DE-dependent parameters from the DE-independent ones.

As peculiar velocities distort cosmological redshifts, when we work in the observed redshift space (denoted by $s$), the correlation function is usually expanded in spherical harmonics. Here, we consider the first term of the expansion, which is also the largest, that is the correlation-function monopole $\xi_{0}$\footnote{Extension to the correlation function quadrupole that will allow us to remove degeneracies in the parameter estimation is left for future work.}. We assume $\xi_{0}$ is given by the following analytic approximation:
\bea \label{peak:nl:s}
	\xi_{0}(s) &=& \frac{1}{4\pi^2} \int_{-1}^{1} \di \mu \int \di\ln{k}\,\,k^{3}P_{\rm lin}(k, z)  \nn \\
	 &&\times~~ \left[b_{10}+b_{01}k^{2}+\mu^{2}f\right]^{2}  \Sigma^{{\rm gal}}(k,\mu)  \\
	 &&\times~~ e^{- k^{2} \sigma_{v}^{2}(1+\mu^{2}f(2+f))}\, j_{0}(ks) \,. \nn
\eea
where: 
$\mu=\hat{\bk}\cdot\hat{\bz}$ is the cosine of the angle between the line of sight $\hat{\bz}$ and the wave vector ${\bk}$;
$P_{\rm lin}(k,z)$ is the linear matter power spectrum (PS) at redshift $z$;
$b_{10}$ is the Eulerian linear bias and $b_{01}$ is the scale-dependent bias;
$f=\di \ln D/\di \ln a$ is the growth rate;
$j_{0}(x)=\sin(x)/x$ is the zero-order spherical Bessel function;
$\sigma_{v}$ is the one-dimensional dark-matter velocity dispersion in linear theory, given by\be
	\sigma_{v}^{2}(z)=\frac{1}{3} \int \frac{\di^{3} q}{(2 \pi)^{3}}\frac{ P_{\rm lin}(q,z)}{q^{2}} \, ; 
	\label{sigma:v}
\ee  
and the distance dispersion is
\be
	\Sigma^{{\rm gal}}(k,\mu) = \frac{1}{1+k^{2}\mu^{2}\sigma_{r,p}^{2}/2}\, ,
	\label{gal}
\ee
with $\sigma_{r,p}=\sigma_{p}(1+z)/H(z)$, 
for a  physical velocity dispersion $\sigma_{p}$ of galaxies \citep{1994MNRAS.267.1020P, 1996MNRAS.279.1310S, 2013MNRAS.430.2446W}.
In \cite{2016MNRAS.455.2474A} the galaxy-dispersion term $\Sigma^{{\rm gal}}$ was not considered, however we checked that it does not impact the position of the LP in the correlation function. 

Notice that Eq.~(\ref{peak:nl:s}) misses physical effects such as the mode-coupling terms \cite{2009MNRAS.400.1643S, Anselmi:2012cn}, non-local bias terms \cite{2012PhRvD..85h3509C} and velocity bias \cite{2010PhRvD..81b3526D}. Nevertheless, in the BAO range of scales, which we conservatively define to be $60 < s < 130$ Mpc/h, $\xi_{0}(s)$ as described by \eqref{peak:nl:s} reproduces within errors the CF from simulations \cite{Crocce:2005xy, 2007ApJ...664..660E, Matsubara:2007wj, 2015JCAP...07..001P}, and provides a prediction of the LP that is consistent with what is found in simulations at the $\sim 0.5\%$ level \cite{2016MNRAS.455.2474A}. As we will show below, Eq.~(\ref{peak:nl:s}) can be further simplified. We finally notice that, more accurate but still unpublished CF models are currently used in the BAO data analyses \cite{2017MNRAS.464.1640S}. 

At the BAO scales, the scale-dependent bias and the one-dimensional dark-matter velocity dispersion are almost completely degenerate. It is therefore convenient to simplify the CF to the  functional form\footnote{In deriving Eq.~(\ref{nl:xi}), we have neglected the effect of galaxy velocity dispersion, i.e.~we set $\Sigma^{{\rm gal}}(k,\mu)\to1$ and numerically tested that for realistic values of $\sigma_{p}$ \cite{2013MNRAS.430.2446W,2013arXiv1312.4996V}, including such a term causes $\xi_{0}$ to vary by less than $0.5\%$ on BAO scales. %Moreover, the bulk of the effect on the CF induced by such a term can be absorbed by tuning the free exponential factor in Eq.~(\ref{nl:xi}), hence the $\Sigma^{{\rm gal}}(k,\mu)$ contribution can be safely neglected.
}
\bea
	\xi_{0}(s) \simeq \int \frac{\di k}{k} \frac{k^{3}P_{\rm lin}(k,z)}{2 \pi^{2}} A^{2} e^{- k^{2} \sigma_{0}^{2}}\, j_{0}(ks)\,, 
	\label{nl:xi}
\eea
where we have defined
\begin{eqnarray}\label{xi:map}
	A^{2}&=&b_{10}^{2}+\frac{2 b_{10} f}{3}+\frac{f^{2}}{5}, \\
	\sigma_{0}^{2}&=& \frac{\sigma_{v}^2 \left[35 b_{10}^2 \left(f^2+2 f+3\right)+14 b_{10} f \left(3 f^2+6 f+5\right)\right.}{105 \,A^{2}}  \nn \\
	&+&\frac{\left.3 f^2 \left(5 f^2+10 f+7\right)\right]}{105 \,A^{2}} 
		-\frac{2 b_{01}(3 b_{10}+f)}{3\,A^{2}}\,\nn .
	%&+&\quad\frac{\sigma_{p}(35 b_{10}^2 +42 b_{10} f + 15 f^2)}{210 \,A^{2}}\, .
\end{eqnarray} 

In the BAO range of scales such a parametrized template reproduces the CF model given by Eq.~(\ref{peak:nl:s}) at the $1.5\%$ level. This is well within the errors of the CF from N-body simulations used to test the validity of Eq.~(\ref{peak:nl:s}). (See e.g. \cite{2015JCAP...07..001P, 2016MNRAS.455.2474A}.) The template \eqref{nl:xi} recovers the LP position at the $0.15\%$ level. 

\subsubsection{Band-Averaged Synthetic Data and Covariance}

We generate band-averaged CF synthetic data by computing Eq.~(\ref{nl:xi}) in spatial bins, such that the CF in the $i$-th bin of size $\Delta{s}$ and volume $V_{s_i}$ is:
\bea	\label{binavxsi}
	 \bar{\xi}_{0}(s_{i}) &=& \frac{1}{V_{s_i} } \int_{V_{s_i}} \di^{3}s \; \xi_{0}(s)  \\
	 &=&  \int \frac{\di k}{k} \frac{k^{3}P_{\rm lin}(k,z)}{2 \pi^{2}} A^{2} e^{- k^{2} \sigma_{0}^{2}}\, \bar{{j}_0}(ks_i)\, , \nn
\eea
where 
\bea
	\label{Vsiinbin}
	V_{s_i}=4\pi s_i^2\Delta s \left[1+\frac{1}{12}
	\left(\frac{\Delta s}{s_i}\right)^2\right] \,,
\eea
and $\bar{{j}_0}(ks_i)$ is the bin-averaged zeroth-order spherical Bessel function 
\be \bar{{j}_0}(ks_i)\equiv 
\frac{\left.s^2j_1(ks)\right|_{s_1}^{s_2}}
{s_i^2k\Delta s\left[1+\frac{1}{12}\left(\frac{\Delta s}{s_i}\right)^2\right]} \ \ , \ \
\left\{\begin{array}{l}
s_2 = s_i+\Delta s/2 \\
s_1 = s_i-\Delta s/2\end{array}\right. ,
\ee
with $j_{1}(x)=\sin(x)/x^{2} -\cos(x)/x$ the first-order spherical Bessel function.

%%%%%%%%%%%%%%%%%%%%%%%%%%%%%%%%%%%%%%%%%%%%%%%%%%%%%%%%%%%%%%%%%%%%%%%%%%%%%%%%%%%%

We work in the small bin-size approximation ($\Delta s/s\ll 1$) and focus on large scales ($s \sim 100 $ Mpc/h). In this regime, the density field is approximately Gaussian \citep{2008PhRvD..77d3525S}, and following \citep{2009MNRAS.400..851S,2016MNRAS.457.1577G} we compute the CF covariance as 
\be
	 C_{ij} = \frac{1}{V_{\mu}} \int \frac{\di k\, k^{2}}{2 \pi^{2} } \bar{j_{0}}(k s_{i})\bar{j_{0}}(k s_{j})\sigma^{2}_{P}(k) \, ,
	\label{cov}
\ee
where $V_{\mu}$ represents the survey volume. $
\sigma_{P}^{2}$ is the Gaussian-plus-Poisson (i.e. discrete) variance 
of the linear power spectrum in redshift space:
\be
	 \sigma_{P}^{2}(k)=\int_{-1}^{1} \di \mu \left[ (b_{10}+f \mu^{2} )^{2}P_{\rm lin}(k) + \frac{1}{\bar{n}_{g}}  \right]^{2}\,,
	\label{sigmaPS}
\ee
where $\bar{n}_{g}$ is the mean number density of galaxies in the survey. 

Notice that we neglect the non-linear covariance corrections, a consistent approximation for the CF in this range of scales \cite{2016MNRAS.457.1577G}. 

%%%%%%%%%%%%%%%%%%%%%%%%%%%%%%%%%%%%%%%%%%%%%%%%%%%%%%%%%%%%%%%%%%%%%%%%%%%%%%%%%%%%
%%%%%%%%%%%%%%%%%%%%%%%%%%%%%%%%%%%%%%%%%%%%%%%%%%%%%%%%%%%%%%%%%%%%%%%%%%%%%%%%%%%%

\subsection{Alcock-Paczynski Distortions and Isotropic Volume Distance}
\label{sec:AP}
In practice, we do not measure comoving positions but rather angles and redshifts. In order to work in comoving coordinates usually a fiducial cosmology is used to convert the measured angles and redshifts to fiducial comoving coordinates. As a consequence, both the comoving coordinates and the correlation function are distorted with respect to the true ones by the so-called Alcock-Paczynski effect. The relation between the distorted (``fid'') and the ``true'' CF monopole can be computed analytically \cite{2013MNRAS.431.2834X}\footnote{We neglect the quadrupole correction, a good approximation if the true cosmology is close enough to the fiducial cosmology as it is always the case in BAO analyses (e.g.~\cite{2016MNRAS.457.1770C}).},
\bea
	\xi_{0}^{{\rm fid}}(s^{\rm fid})\simeq \xi_{0}^{\rm true}\left(D_{V}^{\rm true}(\bar{z})\frac{s^{\rm fid}}{D_{V}^{\rm fid}(\bar{z})}\right) \equiv \tilde{\xi}_{0}^{\rm true}\left(y^{\rm fid}(\bar{z})\right)\, . \nn \\ 
\label{xiAP}
\eea
We have defined the reduced correlation function (RCF) 
$\tilde{\xi}_{0}^{{\rm x}} (t)\equiv \xi_{0}^{{\rm x}} ( D_{V}^{{\rm x}}(\bar{z}) t)$ 
and 
$y^{{\rm x}}(\bar{z}) \equiv s^{{\rm x}}/D_V^{{\rm x}}(\bar{z})$, 
with ${{\rm x}}=\{{{\rm true}},\, {{\rm fid}}\}$ 
and the isotropic volume distance
\be
	D_{V}(\bar{z}) \equiv \left[(1+\bar{z})^{2}D_{A}(\bar{z})^{2}\frac{c\bar{z}}{H(\bar{z})}\right]^{{1/3}}\,,
	\label{xi:y}
\ee
where $D_{A}(z)$ is the angular-diameter distance and $H(z)$ is the Hubble rate.

Corrections to Eq.~(\ref{xiAP}) are negligible on BAO scales, provided that the fiducial value of the mean cosmic matter density is sufficiently close to the true value. 

From Eq.~(\ref{xiAP}) it follows that
\be
	\tilde{\xi}_{0}^{{\rm fid}}(y^{\rm fid}(\bar{z})) \simeq \tilde{\xi}_{0}^{{\rm true}}(y^{\rm fid}(\bar{z}))\, .
\label{xiy}
\ee
Since Eq.~(\ref{xiy}) holds for any 
value of $y^{\rm fid}(\bar{z})$, the functions $\tilde{\xi}_{0}^{{\rm fid}}$ and $\tilde{\xi}_{0}^{{\rm true}}$ are equal,
\be
	\tilde{\xi}_{0}^{{\rm fid}}(y) \simeq \tilde{\xi}_{0}^{{\rm true}}(y)\, .
\label{xianyy}
\ee
Thus if $y_1$ is the location of any feature of the fiducial RCF,
then it is also that feature's location in the true RCF (as usual for a fiducial cosmology sufficiently close to the true one).
The corresponding locations of the features of $\xi_0^{{\rm true}}$ at $s_1^{{\rm true}}$ and $\xi_0^{{\rm fid}}$ at $s_1^{{\rm fid}}$
are therefore related by
\be  
\frac{s_1^{{\rm true}}}{D_V^{{\rm true}}}
\simeq
\frac{s_1^{{\rm fid}}}{D_V^{{\rm fid}}} \, .
\label{feature}
\ee

Henceforth we shall drop the superscript ``true''. Whenever we do not use the ``fid'' superscript ,``true'' is understood.

%%%%%%%%%%%%%%%%%%%%%%%%%%%%%%%%%%%%%%%%%%%%%%%%%%%%%%%%%%%%%%%%%%%%%%%%%%%%%%%%%%%%
%%%%%%%%%%%%%%%%%%%%%%%%%%%%%%%%%%%%%%%%%%%%%%%%%%%%%%%%%%%%%%%%%%%%%%%%%%%%%%%%%%%%

\subsection{Linear Point Cosmic Distance Inference}
\subsubsection{Linear Point Estimation}
In this section we describe the LP-based procedure for cosmic distance estimation, and explain how it provides us a PG-BAO measure. Two LP properties are crucial: 1) the LP is a linear feature; 2) it can be estimated in a model-independent way from CF data.

We first apply Eq.~(\ref{feature}) at the linear point, obtaining
\be
y_{LP}^{\rm fid}(\bar{z})\simeq \frac{s_{LP}}{D_V(\bar{z})}\, ,
\label{yLP}
\ee 
where $s_{LP}$ is unaffected by non-linear effects and therefore redshift-independent, i.e.~insensitive to the underlying dark-energy component of the Universe and to the spatial curvature of the Universe.

We measure $y_{LP}^{\rm fid}(\bar{z})$ by fitting the galaxy CF data with a polynomial function of order five (see \cite{2018PhRvD..98b3527A} for the detailed validation process of the polynomial estimator with mock galaxy data):
\be
\xi_{0}^{\rm fit}(y)=\sum_{i=0}^{5}a_i y^i.
\label{poly}
\ee 
We compute the solutions of $d\xi_{0}^{\rm fit}/dy=0$ to find the location of the peak ($\hat{y}^{\rm fid}_{\rm peak}$) and dip ($\hat{y}^{\rm fid}_{\rm dip}$) in the CF, and obtain an estimate of the location of the LP in dimensionless units by computing the mid-point
\be
\hat{y}^{\rm fid}_{LP}=\frac{1}{2}(\hat{y}^{\rm fid}_{\rm peak}+\hat{y}^{\rm fid}_{\rm dip})\,.
\ee
Since this is given in terms of a combination of the polynomial fitting coefficients, by propagating their errors we obtain the uncertainty on the LP estimate. 

One might worry that although the LP is insensitive to non-linearities in $\Lambda$CDM, 
it might be impacted by non-linearities for non-$\Lambda$CDM cosmologies. In this regard, note that the LP features discovered in \cite{2016MNRAS.455.2474A} hold both for $\Lambda$CDM and smooth-dark-energy models, such as standard and clustering quintessence. In fact, perturbation theory arguments show that the nonlinear propagator, and so the non-linear CF, have the same functional form in all these cases \cite{Anselmi:2012cn, Anselmi:2014nya}. Therefore the linear point remains unaffected by non-linearities, and the model-independent parametric fit, Eq.~(\ref{poly}), applies. We thus conclude that the linear point provides us with a PG-BAO distance measure.

\subsubsection{Optimal Fitting Setup} 

In order to find the optimal {\it setup} for LP fitting, we follow the procedure explained in great detail in \cite{2018PhRvD..98b3527A}. First, we employ the CF model and the covariance matrix (section \ref{sec:synthetic}) to generate $1000$ different synthetic data realizations of the CF\footnote{In \cite{2018PhRvD..98b3527A}, they found that the distribution of the BAO mock CFs is always well-described by a Gaussian. Here, we consistently assume that the CF is Gaussian-distributed, and that Eq.~(\ref{nl:xi}) provides us the mean CF. Therefore, we have all the ingredients to generate the mock realizations.}. Second, from the polynomial fit we require that the distribution of the $\chi^{2}_{{\rm min}}$ is consistent with a $\chi^{2}$ distribution, the mean LP from the mocks is unbiased, the LP statistical error is the smallest possible. We verified that the distribution of the estimated LP values is always consistent with a Gaussian. This provides us the optimal range-of-scale over which to fit the polynomial.

%%%%%%%%%%%%%%%%%%%%%%%%%%%%%%%%%%%%%%%%%%%%%%%%%%%%%%%%%%%%%%%%%%%%%%%%%%%%%%%%%%%%
%%%%%%%%%%%%%%%%%%%%%%%%%%%%%%%%%%%%%%%%%%%%%%%%%%%%%%%%%%%%%%%%%%%%%%%%%%%%%%%%%%%%
\subsection{Correlation Function Model Fitting Cosmic Distance Estimation}
\label{sec:FullFit}  

In this section, cosmic distances are inferred from CF-MF in the BAO range-of-scales. We consider the theoretically motivated model described in Section~\ref{sec:CFmodel} that accounts for the non-linear modifications of the CF with respect to the linear-theory prediction and given by Eq.~(\ref{nl:xi}). To perform a PG-BAO inference, we need to marginalise over the redshift-dependent quantities, which depend on the assumed DE or $\Lambda$CDM model and on the curvature of the Universe.

The Fisher matrix is given by
\be
	 F_{\mu\nu}=\sum_{i, j} \frac{\partial \overline{\tilde{\xi}_{0}}(y^{\rm fid}_{i})}{\partial \theta_{\mu}} C^{-1}_{ij} \frac{\partial \overline{\tilde{\xi}_{0}}(y^{\rm fid}_{j})}{\partial \theta_{\nu}}\, , %+ \frac{1}{2}\text{Tr}\left[ C^{-1} \frac{\partial C}{\partial \theta_{\mu}} C^{-1} \frac{\partial C}{\partial \theta_{\nu}} \right]\,,
	\label{Fisher}
\ee  
where: the Alcock-Paczynski effects are included in $\tilde{\xi}_{0}$, as defined in Eq.~(\ref{xiAP}); $\theta_{\mu}$ is the vector of  model parameters;  $C_{ij}$ is the covariance matrix of $\tilde{\xi}_{0}$ between the $i$-th and $j$-th bins; and $\overline{\bar{\xi}_{0}}(y_{i})$ is the mean value of the RCF in the $i$-th bin\footnote{As usual in current BAO analyses (e.g~\cite{2017MNRAS.470.2617A}), we neglect the parameter dependence of the covariance matrix, both for the LP and the CF-MF approach.}.

We treat~$A$ and $\sigma_{0}$ as free parameters encoding the galaxy-matter bias and non-linear effects. Indeed, the amplitude of density fluctuations, the strength of the non-linear damping, and the bias parameters depend on the DE model \cite{Anselmi:2012cn, Anselmi:2014nya}, the particular galaxy population considered, and the spatial geometry of the Universe. Here, by treating $D_{V}$ as a free parameter, we are able to indirectly account for such dependencies. We are finally left\footnote{For simplicity, in the Fisher-matrix analysis, the fiducial value for $D_{V}$ is chosen to be equal to $D_{V}^{\rm fid}$; however a different value can be chosen, as the fiducial background cosmology is decoupled from the fiducial cosmology for the perturbations.} with $\theta_{\mu}=\{\omega_b,\omega_c,n_s,A,\sigma_0,D_{V}(\bar{z})\}$.
%\subsubsection{Comoving Distance Units}
Note that we do not include $h$ among the parameters we fit. In fact, if the linear CF is expressed as a function of comoving fiducial distances in Mpc units, the $h$ parameter is completely degenerate with $\sigma_{8}$ and the linear local bias $b_{10}$, i.e. it does not affect the shape of the CF. In contrast, in the widely employed Mpc/h units, $h$ spuriously changes the shape of the CF. Non-linearities partially break the degeneracy (see Eqs.~(\ref{nl:xi}) and (\ref{xi:map})) but not for PG-BAO investigations, as the damping parameter is left free. In any case, we discourage use of the widely employed Mpc/h units in clustering analysis, as that induces artificial cosmological parameter dependencies into the observables. 

It is worth noticing that, fitting the CF-model parameters specified by $\theta_{\mu}$, we can infer constraints on the sound-horizon scale $r_d(\omega_b,\omega_c)$, which can be computed with high accuracy by means of the CAMB code \cite{2000ApJ...538..473L, Thepsuriya:2014zda}. This standard-ruler scale is embedded in the CF, and can be consistently estimated through the CF-MF described here. In the next section we will show that the ratio of the CF-MF-inferred sound-horizon to the isotropic volume distance, $r_{d}/D_{V}(\bar{z})$, provides an accurate distance estimator.

%%%%%%%%%%%%%%%%%%%%%%%%%%%%%%%%%%%%%%%%%%%%%%%%%%%%%%%%%%%%%%%%%%%%%%%%%%%%%%%%%%%%
%%%%%%%%%%%%%%%%%%%%%%%%%%%%%%%%%%%%%%%%%%%%%%%%%%%%%%%%%%%%%%%%%%%%%%%%%%%%%%%%%%%%
\section{Results}
\label{sec:results}

We present the forecast errors for $s_{LP}/D_{V}(\bar{z})$ from the LP standard ruler and for $D_{V}(\bar{z})$ and $r_{d}/D_{V}(\bar{z})$ from the CF-MF approach. 

In the following we generate synthetic data for a fiducial $\Lambda$CDM model with cosmological parameter values consistent with Planck data analysis \cite{2016A&A...594A..13P}: $\Omega_{b} = 0.0486$, $\Omega_{c}= 0.259$, $H_{0} = 67.74$, $n_{s}=0.9667$ and $\sigma_{8}=0.831$. 

We consider two realistic galaxy survey configurations corresponding to the DESI mission for low-redshift clustering measurements and the ESA Euclid satellite for high-redshift measurements:

\begin{itemize} 
\item {\bf DESI} \\
The DESI survey is designed as described in \cite{2016arXiv161100036D}. We consider only the low-redshift galaxies target, the Bright Galaxies Surveys (BGSs) covering a redshift range of $0.0 \le z \le 0.5$. Following \cite{2016arXiv161100036D}, we assume a survey sky fraction $f_{{\rm sky}}=0.339$. The forecast number of galaxies per unit redshift per square degree is reported in Tables 2.3 and 2.5 in \cite{2016arXiv161100036D}. The redshift dependence of the linear bias is provided in terms of the growth factor $D(z)$ normalized to $D(z=0)\equiv 1$: 
\bea
	&&b^{BGS}_{10}(z)\, D(z)=1.34 \nn \\
\eea

\item {\bf Euclid} \\
We design our Euclid-like survey following the survey specifications presented in \cite{2011arXiv1110.3193L}
. We assume a sky fraction of $f_{{\rm sky}}=0.375$, a redshift range of $0.6 \le z \le 2.1$ and a total  of 50 million galaxies with the redshift distribution reported in \cite{2011arXiv1110.3193L}\footnote{Notice that there is some uncertainty in the expected number of galaxies. The number used here might yet prove to be optimistic \cite{2014JCAP...05..023F}.}. Following \cite{2016arXiv160600180A}, we model the redshift dependence of the linear bias as 
\be 
	b_{10}(z)=\sqrt{1+z}\, .
\ee

\end{itemize}

There are no  forecasts available of the scale-dependent bias for the BGSs and ELGs, we thus set their fiducial values to zero. We neglect spectroscopic redshift errors.

We consider redshift bins large enough to detect the BAO signal also in the transverse direction $\Delta z  = 0.2$, and for the first redshift bin we choose $\Delta z  = 0.3$ in order to have a sufficiently large volume to detect the BAO. For the DESI survey we focus on redshift bins in the range $0.0 \le z \le 0.5$; for Euclid we consider $0.6 \le z \le 2.0$. For each bin, we use the simplified assumption that the effective redshift is given by the mean redshift. This simplification can be easily dropped without any impact on our message.

To be concise we will discuss in detail the results for the lowest Euclid redshift bin and simply quote the results for all the other redshift bins.

\subsection{Linear Point standard ruler: $s_{LP}/D_{V}(\bar{z})$}
We apply the methodology explained in \cite{2018PhRvD..98b3527A} to determine the LP optimal fitting setup. We find that the quintic polynomial must be fit over the range of scales $75 < s < 115$ Mpc/h for all the redshift bins except for $\bar{z}=0.4$ for which the scale range is $70 < s < 120$ Mpc/h. The bin width is $\Delta s = 3$ Mpc/h. With this setup, for the lowest Euclid redshift bin, the mean of the distance errors is
\be
\overline{\left(\sigma_{\frac{s_{LP}}{D_{V}(\bar{z})}}/\frac{s_{LP}}{D_{V}(\bar{z})}\right)} \times 100 = 1.1\%\, .
\ee 

 %%%%%%%%%%%%%%%%%%%%%%%%%%%%%%%%%%%%%%%%%%%%%%%%%%%%%%%%%%%%%%%%%%%%%%%%%%%%%%%%%%%%
%%%%%%%%%%%%%%%%%%%%%%%%%%%%%%%%%%%%%%%%%%%%%%%%%%%%%%%%%%%%%%%%%%%%%%%%%%%%%%%%%%%%
\subsection{Correlation Function Model Fitting: $r_{d}/D_{V}(\bar{z})$}
\label{sec:results:model}
Employing the Fisher matrix formalism for the CF-MF approach (see Section \ref{sec:FullFit}) we forecast, for the lowest Euclid redshift bin, the following marginalized errors for the six fit parameters $\theta_{\mu}=\{\omega_b,\omega_c,n_s,A,\sigma_0,D_{V}(\bar{z})\}$: 
\be
(\sigma_{\theta_{\mu}}/\theta_{\mu})\times 100=\{230, 150, 45, 50, 90, 72\}\%\, .
\ee 

As already mentioned, a change of variables to $\tilde{\theta}_{\mu}=\{r_{d}(\omega_{b},\omega_{c}),\omega_c,n_s,A,\sigma_0,D_{V}(\bar{z})\}$ returns
\be
(\sigma_{\tilde{\theta}_{\mu}}/\tilde{\theta}_{\mu})\times 100=\{70, 150, 45, 50, 90, 72\}\%\, ,
\ee 
and a Pearson correlation coefficient between $r_{d}$ and $D_{V}(\bar{z})$  of
\be
\rho_{r_{d},D_{V}}=0.99992\, .
\ee 
Since the $r_{d}$ and $D_{V}(\bar{z})$ relative errors differ by a small amount, and the Pearson correlation coefficient is very close to unity, we conclude that the isotropic-volume distance is almost completely determined by the sound-horizon scale. Hence, even if the individual measurements of $r_{d}$ and $D_{V}(\bar{z})$ are not measured well, their ratio is measured with very good precision and it yields
\be
\left(\sigma_{\frac{r_{d}}{D_{V}(\bar{z})}}/\frac{r_{d}}{D_{V}(\bar{z})}\right) \times 100 = 1.8\%\, .
\ee 

The numerical CAMB code that we employ to perform an accurate Fisher matrix analysis introduce numerical instabilities that need to be mitigated. The methodology that we develop in this regard is detailed in Appendix \ref{appendix:Fisher}.

\begin{table}
\vspace{1cm}
\renewcommand{\arraystretch}{1.1}
\begin{center}
\textbf{Linear Point and CF-MF distance errors} \\
\vspace{0.3cm}
\begin{scriptsize}
\begin{tabular}{| C{0.5cm} | C{1.8cm} | C{1.2cm} | C{1.1cm} C{1.2cm} C{1.7cm}  |} 
\hhline{~|-|----} 
\multicolumn{1}{c|}{} & \multicolumn{1}{c|}{}& \multicolumn{4}{c|}{}  \\
\multicolumn{1}{c|}{} & \multicolumn{1}{c|}{\textbf{LP}}& \multicolumn{4}{c|}{ \textbf{CF-MF}}  \\ [0.2cm]
%\hhline{-|~|~~~~} 
\hline 
&&&&& \\
$\bar{z}$ & $\frac{s_{LP}}{D_{V}(\bar{z})}$  & $\frac{r_{d}}{D_{V}(\bar{z})}$ & $r_{d}$  & $D_{V}(\bar{z})$ & $\rho_{r_{d},D_{V}}$ \\  
&&&&& \\
\hline 
%\multicolumn{1}{|c|}{} &\multicolumn{6}{c|}{ \textbf{DESI-like}}  \\ [0.2cm]
%\hline 
&&&&& \\
$0.15$  & $\mathbf{2.9\%}$ & $\mathbf{6.1}\%$& 218.1\%   & 223.5\% & 0.99992 \\ [0.2cm]
$0.4$    & $\mathbf{2.3\%}$ & $\mathbf{4.3\%}$& 146.8\%   & 150.6\% & 0.99992 \\ [0.2cm]
\hline 
&&&&& \\
$0.7$ & $\mathbf{1.2\%}$ & $\mathbf{1.8\%}$& 70.0\%   & 71.6\% & 0.99992 \\ [0.2cm]
$0.9$ & $\mathbf{0.9\%}$ & $\mathbf{1.3\%}$& 52.6\%   & 53.7\% & 0.99992 \\ [0.2cm]
$1.1$ & $\mathbf{0.7\%}$ & $\mathbf{1.1\%}$& 45.6\%   & 46.5\% & 0.99992 \\ [0.2cm]
$1.3$ & $\mathbf{0.7\%}$ & $\mathbf{1.0\%}$& 42.5\%   & 43.3\% & 0.99991 \\ [0.2cm]
$1.5$ & $\mathbf{0.8\%}$ & $\mathbf{1.0\%}$& 43.8\%   & 44.6\% & 0.99991 \\ [0.2cm]
$1.7$ &$\mathbf{1.1\%}$ & $\mathbf{1.3\%}$& 52.7\%   & 53.8\% & 0.99990 \\ [0.2cm]
$1.9$ &$\mathbf{2.0\%}$ & $\mathbf{2.3\%}$& 83.1\%   & 85.1\% & 0.99990 \\ [0.3cm]
\hline
\end{tabular}
\end{scriptsize}
\end{center}
\caption[]{\label{tab:LPrd} 
We show the forecasted precision of distance measurements for the LP standard ruler and CF-MF approach. The first two raws represent the DESI low redshift bins, the following rows the Euclid bins. Notice that all the shown digits are necessary to compute the $r_{d}/D_{V}(\bar{z})$ error.}
\end{table}

 %%%%%%%%%%%%%%%%%%%%%%%%%%%%%%%%%%%%%%%%%%%%%%%%%%%%%%%%%%%%%%%%%%%%%%%%%%%%%%%%%%%%
%%%%%%%%%%%%%%%%%%%%%%%%%%%%%%%%%%%%%%%%%%%%%%%%%%%%%%%%%%%%%%%%%%%%%%%%%%%%%%%%%%%%
\subsection{Discussion}
In Table \ref{tab:LPrd}, we present the PG-BAO-forecasted distance errors from the LP standard ruler and the CF-MF approach. Given the survey characteristics considered here, we find that the LP provides up to $50\%$ smaller statistical errors than those on the distance estimates from the CF-MF method. Notice that both $s_{LP}$ and $r_{d}$ depend only on the $\omega_{b}$ and $\omega_{c}$ parameters \cite{2016MNRAS.455.2474A}. Moreover, when derived from Plank CMB measurements, their relative errors are identical \cite{LP_CMB}. Therefore, using estimates of $s_{LP}/D_{V}$ or $r_{d}/D_{V}$ for cosmological parameter constraints purposes is entirely equivalent, it is only the accuracy and the precision with which the ratio is measured that matters\footnote{Here we ignore the $0.5\%$ intrinsic uncertainty of the LP and assume that the CF model is unbiased.}. 

Both the CF-MF method and the LP standard ruler polynomial estimator need to be 
validated using N-body simulations and mock galaxy catalogues. This will provide us with the identification of the optimal range of scales for the CF fits. In this work, we cannot apply this procedure to the CF-MF approach, since the model fitting function coincides with the one we use to generate our synthetic CF data. Hence, we have simply checked that extending the analysis by 40 Mpc/h (see Section \ref{sec:CFmodel}) reduces the $r_{d}/D_{V}$ error only by $\sim 5\%$ which does not impact our conclusions.

The reader might be concerned that, given the large degeneracies among parameters, a Fisher-matrix approach might yield a poor approximation of the true error. However, the Cramer-Rao bound holds and implies that 
\be
	\sigma^{{\rm Fisher}}_{r_{d}/D_{V}(\bar{z})} \le \sigma_{r_{d}/D_{V}(\bar{z})}\, .
\ee 
This is strictly true if, instead of using $\tilde{\theta}_{\mu}=$ $\{r_{d}(\omega_{b},\omega_{c}),$ $\omega_c,n_s,A,\sigma_0,D_{V}(\bar{z})\}$ as in Section \ref{sec:results:model} we consider $\tilde{\theta}_{\mu}=\{r_{d}(\omega_{b},\omega_{c})/D_{V}^{\rm true}(\bar{z}),\omega_c,n_s,A,\sigma_0,D_{V}^{\rm true}(\bar{z})\}$. We have verified that both computations provide exactly the same results. Therefore the differences between the distance errors from the LP standard ruler and CF-MF approach might be larger or equal to those presented in Table \ref{tab:LPrd}. However, they will not change sign, thus leaving our main conclusions unaltered.

A more comprehensive treatment than the one presented here could be achieved by running Monte-Carlo-Markov-Chains (MCMCs) on synthetic mocks. This would allow one to estimate the ``detection probability'' for the distance estimates. In fact, given the finite volume of the surveys, the BAO feature in the CF might not be present due to cosmic variance. In the LP case, this would manifest with the polynomial estimator previously defined not ``detecting'' the peak and dip in the BAO range of scales, i.e.~$d\xi_{0}^{\rm fit}/dy=0$ would not have real solutions. For the CF-MF approach, the ratio of the number of mocks for which the chains have converged to the total number of mocks would provide an estimate of the detection probability. Moreover a MCMC analysis would improve on the local expansion-based method that we use to estimate the LP error. Finally, MCMCs allow to properly estimate the $r_{d}/D_{V}$ error that is equal or larger than the one we computed with the Fisher matrix formalism. We defer such a computationally involved analysis to more realistic mocks and more accurate CF models.

\subsubsection{Testing the BAO-Only error estimation}
As explained in the introduction, the standard practice to extract geometric distance information from the CF is the BAO-Only methodology (e.g~\cite{2008ApJ...686...13S, 2012MNRAS.427.2146X, 2014MNRAS.441...24A}): the fit is performed by keeping fixed to the fiducial value the cosmological and damping parameters, while broad-band (BB) terms are added to Eq.~(\ref{nl:xi}). The new BB piece reads
\be
	\tilde{\xi}^{{\rm BB}}_{0}(y^{\rm fid})\equiv \frac{a_{1}}{(D_V^{{\rm fid}}  y^{\rm fid})^{2}} + \frac{a_{2}}{D_V^{{\rm fid}}  y^{\rm fid}} + a_{3} \, .
\ee 
This new contribution should take into account theoretical non-linear distortions, mitigate the impact of fixing the fundamental parameters and make the fit insensitive to observational systematic effects \cite{2017MNRAS.464.1168R}. With the BAO-O we thus fit 5 parameters: $\theta_{\mu}=\{a1, a2, a3, A,D_{V}(\bar{z})\}$. As explained in Section 7 of \cite{2008ApJ...686...13S}, fixing the cosmological parameters implies keeping $r_{d}$ fixed in the CF fit; consequently we need to reinterpret the estimated $D_{V}(\bar{z})$ value and error as $\alpha_{y}\equiv D_{V}(\bar{z})/(r_{d}/r^{\rm fid}_{d})$ \footnote{Note that in the standard notation, where Eq.~(\ref{xiAP}) is written in terms of $s^{\rm fid}$ and not $y^{\rm fid}$, $\alpha_{y}$ corresponds to $\alpha\equiv (D_{V}(\bar{z})/D_{V}^{\rm fid}(\bar{z}))/(r_{d}/r^{\rm fid}_{d})$.}.

We test, by means of the Fisher matrix, whether the addition of the broad-band terms precisely recovers the uncertainties neglected by the fixing of the cosmological and damping parameters\footnote{In the fitting method presented in \cite{2012MNRAS.427.2146X}, a slightly different model than Eq.~(\ref{peak:nl:s}) is used. We checked that they have the same cosmological and damping parameters, and they agree at ~4\% at BAO scales. Since there are no fundamental reasons to prefer one of the two models, we employ our synthetic-data model, which is used in more recent theoretical investigations \cite{2015JCAP...07..001P, 2015PhRvD..92d3514B}.}. In Table \ref{tab:fixing}, we compare $\sigma_{r_{d}/D_{V}(\bar{z})}$ estimated from the approach of Section \ref{sec:results:model} to the BAO-O values. The numerical results are presented in Table \ref{tab:fixing} and show that, by fixing the cosmological and the damping parameters, we are underestimating the distance-measurement statistical uncertainty by nearly a factor of $2$. We have verified that the result does not change if we allow the damping parameter to vary in the BAO-O fit. 

In light of these results, we argue that the BAO-O fitting procedure, in which the cosmological and the damping parameters are kept fixed and broad-band terms are added, should be reconsidered. 

\begin{table}
\vspace{1cm}
\renewcommand{\arraystretch}{1.1}
\begin{center}
\textbf{Testing the BAO-Only error estimation} \\
\vspace{0.3cm}
\begin{scriptsize}
\begin{tabular}{| C{0.5cm} | C{3cm}  | C{3cm} |} 
\hhline{~|-|-} 
\multicolumn{1}{c|}{} & \multicolumn{1}{c|}{} & \multicolumn{1}{c|}{} \\
\multicolumn{1}{c|}{} & \multicolumn{1}{c|}{\textbf{CF-MF}}  & \multicolumn{1}{c|}{\textbf{BAO-Only}} \\ [0.2cm]
\multicolumn{1}{c|}{} & \multicolumn{1}{c|}{\textbf{}}  & \multicolumn{1}{c|}{\textbf{}} \\ [0.2cm]
%\hhline{-|~|~~~~} 
\hline 
&& \\
$\bar{z}$ & $\frac{r_{d}}{D_{V}(\bar{z})}$  & $\frac{r_{d}}{D_{V}(\bar{z})}$  \\  
&& \\
\hline 
&& \\
$0.15$  & $\mathbf{6.1\%}$ & $\mathbf{2.7}\%$ \\ [0.2cm]
$0.4$  & $\mathbf{4.3\%}$ & $\mathbf{2.0}\%$ \\ [0.2cm]
\hline
&& \\
$0.7$  & $\mathbf{1.8\%}$ & $\mathbf{0.9}\%$ \\ [0.2cm]
$0.9$  & $\mathbf{1.3\%}$ & $\mathbf{0.7}\%$ \\ [0.2cm]
$1.1$  & $\mathbf{1.1\%}$ & $\mathbf{0.6}\%$ \\ [0.2cm]
$1.3$  & $\mathbf{1.0\%}$ & $\mathbf{0.6}\%$ \\ [0.2cm]
$1.5$  & $\mathbf{1.0\%}$ & $\mathbf{0.6}\%$ \\ [0.2cm]
$1.7$  & $\mathbf{1.3\%}$ & $\mathbf{0.8}\%$ \\ [0.2cm]
$1.9$  & $\mathbf{2.3\%}$ & $\mathbf{1.4}\%$ \\ [0.2cm]
\hline
\end{tabular}
\end{scriptsize}
\end{center}
\caption[]{\label{tab:fixing} 
We show the forecasted precision for distance measurements. We compare the case where all the cosmological parameters are free to vary with the BAO-only procedure where the fundamental parameters are kept fixed in the fit and broad-band terms are added. The first two raws represent the DESI low redshift bins, the following rows the Euclid bins. }
\end{table}

%%%%%%%%%%%%%%%%%%%%%%%%%%%%%%%%%%%%%%%%%%%%%%%%%%%%%%%%%%%%%%%%%%%%%%%%%%%%%%%%%%%%
%%%%%%%%%%%%%%%%%%%%%%%%%%%%%%%%%%%%%%%%%%%%%%%%%%%%%%%%%%%%%%%%%%%%%%%%%%%%%%%%%%%%
\section{Conclusions}
\label{sec:concl}

Baryon Acoustic Oscillations (BAO) are key to mapping the cosmic expansion history of the universe 
\cite{Bassett:2009mm, 2018JCAP...05..033H}, since they provide a cosmological standard ruler to infer model-independent distance measurements. 
However, to date this property has not been rigorously exploited. To clarify this point, we have introduced two Purely-Geometric-BAO approaches. These methods trade
 tracer and model independencies for larger statistical uncertainties on cosmic distance estimates. Their key advantage is therefore that the error estimation does not have to rely on N-body simulations and mock catalogs covering the possibly infinite space of time-varying DE equation-of-state models and parameter values.

 Using a set of synthetic CF data, generated assuming future DESI and Euclid survey configurations, we show practical implementations of these two PG-BAO methods to infer cosmic distance estimates. The first is the Linear Point standard ruler, which we investigate following the analyses presented in \cite{2016MNRAS.455.2474A, 2018PhRvD..98b3527A, 2018PhRvL.121b1302A}. The second is a novel approach that we introduce here for the first time, which relies on fitting a model of the CF to the data, and that allows us
to properly estimate $r_{d}/D_{V}(z)$ while consistently propagating all the CF-model parameter uncertainties. We compare the results of the two implementations. We find that the LP-inferred errors are up to $50\%$ smaller than the CF-MF ones. 

We also test the approximations employed in BAO-Only analyses, as defined in \cite{2008ApJ...686...13S, 2012MNRAS.427.2146X}, in which distance estimates are inferred from CF-template fitting. In this fitting procedure, the cosmological and damping parameters are kept fixed to a fiducial value, while three broad-band parameters are added.  
We find that this approach underestimates by nearly a factor of 2 the cosmic distance uncertainties. Hence, our analysis suggests that cosmological parameter constraints inferred from the BAO-O distance measurements alone or in combination with other cosmological probes might be overly optimistic and potentially biased \cite{2018JCAP...05..033H, 2017NatAs...1..627Z, 2015PhRvD..92l3516A}. We thus suggest to reconsider this methodology. Ideally, a cosmological measurement should be as agnostic as possible of the underlying cosmological model; it is for this  reason that we have introduced the two PG-BAO approaches. 

An equally important result of our work is the formulation of a statistically rigorous BAO forecasting method that improves on the currently used procedure based on \cite{2007ApJ...665...14S}.

The CF model we have assumed throughout the paper, Eq.~(\ref{peak:nl:s}), can be improved. On this point, a number of BAO analyses have used the more accurate gRPT model for the computation of the CF \cite{2017MNRAS.464.1640S}. It would be interesting to implement the CF-MF approach using the gRPT model, carefully identifying all the tracer and redshift-dependent parameters, marginalising over them and comparing the inferred distance estimate against the LP one (which on the other hand does not need to specify a CF fitting model). 

In our treatment we did not consider the observational systematic effects \cite{2017MNRAS.464.1168R}. In BAO analyses, they are usually mitigated by using weights. Another approach is provided by the BAO-Only fits, which marginalise over systematics. We expect the LP to be insensitive to the main BOSS observational distortions that manifest as a constant additive term to the CF \cite{2017MNRAS.464.1168R}. In fact, the LP is estimated through first-order CF spatial derivatives, which are not affected by an additive constant. On the other hand, an extra parameter needs to be added to the CF-MF approach to marginalise over it, probably worsening its distance-constraining power. This might be another advantage of the LP.

Through the CF-MF analysis, we find that $r_{d}/D_{V}(z)$ is measured with exquisite precision. This is given by the sound-horizon scale being closely related to the BAO feature position; however, there is no fundamental reason that forces us to use $r_{d}$. It would thus be interesting to look for the optimal function $r_{x}(\omega_{c},\omega_{b})$ that minimises the statistical error of $r_{x}/D_{V}(z)$.

Here, we have discussed PG-BAO methods for the CF monopole, thus providing measurements of the isotropic-volume-distance. However, it is well known that, in order to break the degeneracy between the Hubble parameter and the angular-diameter distance, it is crucial to extend the CF analysis to the quadrupole term \cite{2013MNRAS.431.2834X}. Properly complementing both the LP standard ruler and the CF-MF approach with the quadrupole analysis is the subject of work in progress. 

Throughout our analysis we always assumed that neutrinos are massless, still a common approximation in most of BAO studies. It is thus crucial to investigate the impact of massive neutrinos on the PG-BAO distance measurements both employing dedicated N-body simulations and analytical CF models \cite{2015JCAP...07..043C, 2016JCAP...07..034C, 2015JCAP...07..001P, Thepsuriya:2014zda}.

In this paper we explain how to implement the CF-MF method. Remarkably, the same procedure can be similarly applied to the power spectrum through PS-model-fitting (PS-MF). The most relevant difference is that the mode-coupling term plays a relevant role for the PS-MF, introducing a different redshift and tracer dependence compared to the CF case. However, this impacts only the practical implementation; there is no difference at a more fundamental level.

The work presented here is intended to make BAO a more rigorous tool for cosmological studies. PG-BAO-based distance measurements can in fact be used to discriminate among specific classes of cosmological models (i.e.~$\Lambda$CDM and smooth standard/clustering quintessence models, without the assumption of spatial flatness). Thus, a relevant step forward would be to extend this analysis to non-smooth quintessence models such as those where dark-matter is non-minimally coupled to dark-energy \cite{2012PDU.....1..162B} or where the sound speed of dark energy is neither equal to the speed of light nor vanishing \cite{Anselmi:2011ef}.

\begin{acknowledgments} 
SA thank Stephane Colombi for clarifying discussions on the meaning of a fit where the fundamental parameters are kept fixed. GDS thanks the IAP and the Observatoire de Paris for their frequent hospitality. GDS was partially supported by a Department of Energy grant DE-SC0009946 to the particle astrophysics theory group at CWRU. 
The research leading to these results has received funding from the European Research Council under the European Community Seventh Framework Programme (FP7/2007-2013 Grant Agreement no. 279954) ERC-StG "EDECS". IZ is supported by National Science Foundation grant AST-1612085.
\end{acknowledgments}

\newpage

\begin{widetext}
\appendix

\section{FISHER MATRIX NUMERICAL STABILITY}
\label{appendix:Fisher}

\subsection{Fisher matrix convergence tests}
\label{appendix:derivatives}
To obtain the covariance matrix for the fitted parameters the Fisher matrix must be inverted, this non-linear operation cause numerical instabilities that we address by performing converge tests as outlined below.

It is convenient to work with the rescaled parameters $\theta_{\mu}^{{\rm R}}\equiv\theta_{\mu}/\theta_{\mu}^{{\rm fid}}$, where $\theta_{\mu}^{{\rm fid}}$ means it is evaluated at the fiducial value. The RCF derivatives that enter in the Fisher matrix equation (\ref{Fisher}) can be written in the following way
\be
	\frac{\partial \overline{\tilde{\xi}_{0}}(y^{\rm fid}_{i})}{\partial \theta_{\mu}^{{\rm R}}} = \int \di k\; \frac{ k^{2}\, \bar{j}_{0}(k\,y^{\rm fid}_{i}\,D_{V})}{2 \pi^{2}} \frac{\partial P(k, z)}{\partial \theta_{\mu}^{{\rm R}}} \, .
	\label{xideriv}
\ee
To be accurate we employ the CAMB code \cite{2000ApJ...538..473L} to compute the power spectrum. However this implies that we cannot use the five-point stencil algorithm to compute the PS numerical derivatives. In fact the CAMB code outputs are affected by numerical noise that needs to be tamed to obtain accurate enough numerical derivatives. Therefore, we evaluate $\partial P(k, z)/\partial \theta_{\mu}^{{\rm R}}$ by means of the following procedure. For each parameter we compute the PS on $L$ points ($L$ is an odd number) equally spaced in $\theta_{\mu}^{{\rm R}}$ around the fiducial value $\theta_{\mu}^{{\rm R}}=1$ and with extrema $\theta_{\mu}^{{\rm R, max}}=1.05$ and $\theta_{\mu}^{{\rm R, min}}=0.95$. For each $k$ value of the CAMB output we fit the PS to a polynomial around $\theta_{\mu}^{{\rm R}}=1$
\be
	P^{{\rm fit}}(k;\, \theta_{\mu}^{{\rm R}})=b_{1}+b_{2}(\theta_{\mu}^{{\rm R}}-1)+b_{3}(\theta_{\mu}^{{\rm R}}-1)^{2} + b_{4}(\theta_{\mu}^{{\rm R}}-1)^{3} + \cdots + b_{N} (\theta_{\mu}^{{\rm R}}-1)^{N-1}\ \, .
	\label{}
\ee
We verify that a cubic polynomial interpolation returns randomly distributed residuals for $P^{{\rm CAMB}}(k;\, \theta_{\mu}^{{\rm R}})-P^{{\rm fit}}(k;\, \theta_{\mu}^{{\rm R}})$ as a function of $\theta_{\mu}^{{\rm R}}$. The $b_{2}$ coefficient provides us the estimate of $(\partial P^{{\rm fit}}(k;\, \theta_{\mu}^{{\rm R}})/\partial \theta_{\mu}^{{\rm R}})\vert_{\theta_{\mu}^{{\rm R}}=1}$. A high value of $L$ reduces the impact of the CAMB noise on the $b_{2}$ evaluation. We test the stability of the parameters' covariance matrix by performing converge tests. We compute $b_{2}$ from three different subsets of $\{\{\theta_{\mu,1}^{{\rm R}},P^{{\rm CAMB}}(k;\, \theta_{\mu,1}^{{\rm R}})\},...,\{\theta_{\mu,L}^{{\rm R}},P^{{\rm CAMB}}(k;\, \theta_{\mu,L}^{{\rm R}})\}\}$: the first set contains all the L data points, the second set consists of $\{\{\theta_{\mu,1}^{{\rm R}},P^{{\rm CAMB}}(k;\, \theta_{\mu,1}^{{\rm R}})\},\{\theta_{\mu,3}^{{\rm R}},P^{{\rm CAMB}}(k;\, \theta_{\mu,3}^{{\rm R}})\},...,\{\theta_{\mu,L}^{{\rm R}},P^{{\rm CAMB}}(k;\, \theta_{\mu,L}^{{\rm R}})\}\}$ and the third is given by $\{\{\theta_{\mu,2}^{{\rm R}},P^{{\rm CAMB}}(k;\, \theta_{\mu,2}^{{\rm R}})\},\{\theta_{\mu,4}^{{\rm R}},P^{{\rm CAMB}}(k;\, \theta_{\mu,4}^{{\rm R}})\},...,\{\theta_{\mu,L-1}^{{\rm R}},P^{{\rm CAMB}}(k;\, \theta_{\mu,L-1}^{{\rm R}})\}\}$. We verified that the covariance matrixes obtained from the second and the third sets of points agree with the first one at the $2\%$ level.

We checked that the bin size adopted for $y_{i}^{{\rm fid}}$ in the computation of $\overline{\tilde{\xi}_{0}}(y^{\rm fid}_{i})$ does not change the Fisher matrix outcomes, as expected.

Finally the same method we employed to reduce the CAMB numerical noise can be applied directly to the RCF derivatives without using the convenient Eq.~(\ref{xideriv}). We perform this computation as a cross-check and find that the covariance matrix differ by $10\%$ with respect to the computation above presented. We do not track the origin of this discrepancy but we notice that it causes an error on the parameters' errors of order $5\%$ which does not impact the conclusions of the analysis presented in the main text.

\subsection{Fisher matrix: change of basis}
\label{appendix:proj}
In Section \ref{sec:results:model} in order to estimate the error on $r_{d}/D_{V}(\bar{z})$ we need to project the Fisher matrix from the standard parameters basis $\theta_{\mu}=\{\omega_b,\omega_c,n_s,A,\sigma_0,D_{V}(\bar{z})\}$ to the following set of parameters $\tilde{\theta}_{\mu}=\{r_{d}(\omega_{b},\omega_{c}),\omega_c,n_s,A,\sigma_0,D_{V}(\bar{z})\}$. The $\tilde{\theta}_{\mu}$ Fisher matrix reads \cite{2010deto.book.....A}
\be
	F^{(\tilde{\theta}_{\mu})}_{lm}=J_{il}F^{(\theta_{\mu})}_{ij}J_{jm} \, ,
	\label{}
\ee
where $J=(\partial \tilde{\theta}_{\mu} / \partial \theta_{\mu})^{-1}$ is the inverse of the Jacobian evaluated at the fiducial parameter values (i.e.~maximum likelihood). To compute $\partial r_{d}(\omega_{b},\omega_{c}) / \partial \theta_{\mu}$ we employ the same polynomial fitting technique explained in Appendix \ref{appendix:derivatives}. We obtain the same level of convergence for the Fisher matrix for $\tilde{\theta}_{\mu}$ as the one for the original parameters. We verified that the numerical systematic uncertainty on the estimated error of $r_{d}/D_{V}(\bar{z})$ is of the order of $2\%$. We finally point out that we need to quote many significant digits in the uncertainties reported in Table \ref{tab:LPrd} even if our numerical accuracy is much smaller, however this is consistent as the systematic differences between covariance matrixes cancel out in the $r_{d}/D_{V}(\bar{z})$ error propagation formula.

\end{widetext}

\bibliography{MyBib}

\end{document}